\documentclass[journal,12pt,draftclsnofoot,onecolumn]{IEEEtran}
\usepackage[noadjust]{cite}

\usepackage[T1]{fontenc}
\usepackage{hyperref}
\usepackage[utf8]{inputenc}
\usepackage{color}
\usepackage{caption}
\usepackage{subfigure}
\usepackage{bbm}
\usepackage{mathtools}
\usepackage{graphicx}
\usepackage{amsmath}
\usepackage[cmintegrals]{newtxmath}
\usepackage{comment}
\usepackage{amssymb}
\usepackage{algorithm}
\usepackage{algpseudocode}
\IEEEoverridecommandlockouts

\newcommand{\icol}[1]{
  \left[\begin{smallmatrix}#1\end{smallmatrix}\right]%
}
\newcommand{\rom}[1]{\uppercase\expandafter{\romannumeral #1\relax}}

\algnewcommand{\LineComment}[1]{\State \(\triangleright\) #1}

\graphicspath{{figures/}}

\newtheorem{theorem}{\textbf{Theorem}}

\DeclareMathOperator*{\argmax}{arg\,max}

\begin{document}

\title{Deep Reinforcement Learning for \\Dynamic Multichannel Access in Wireless Networks}

\author{\small \IEEEauthorblockN{Shangxing Wang,
Hanpeng Liu, Pedro Henrique Gomes and
Bhaskar Krishnamachari}\\
\IEEEauthorblockA{University of Southern California, Los Angeles, USA\\
Email: {shangxiw@usc.edu, lhp13@mails.tsinghua.edu.cn, \{pdasilva, bkrishna\}@usc.edu}}
\thanks{\qquad \qquad This work is an extended version of our ICNC conference paper~\cite{dqn2017}.}}


\maketitle
\begin{abstract}
We consider a dynamic multichannel access problem, where multiple correlated channels follow an unknown joint Markov model. A user at each time slot selects a channel to transmit data and receives a reward based on the success or failure of the transmission. The objective is to find a policy that maximizes the expected long-term reward. 
The problem is formulated as a partially observable Markov decision process (POMDP) with unknown system dynamics. To overcome the challenges of unknown system dynamics as well as prohibitive computation, we apply the concept of reinforcement learning and implement a Deep Q-Network (DQN) that can deal with large state space without any prior knowledge of the system dynamics. 
We provide an analytical study on the optimal policy for fixed-pattern channel switching with known system dynamics and show through simulations that DQN can achieve the same optimal performance without knowing the system statistics. We compare the performance of DQN with a Myopic policy and a Whittle Index-based heuristic through both simulations as well as real-data trace and show that DQN achieves near-optimal performance in more complex situations. Finally, we propose an adaptive DQN approach with the capability to adapt its learning in time-varying, dynamic scenarios.
\end{abstract}


\IEEEpeerreviewmaketitle

\section{Introduction}
\label{sec:introduction}

\IEEEPARstart{P}{rior} work~\cite{knoop, akyildiz2006next} has shown that dynamic spectrum access is one of the keys to improving the spectrum utilization in wireless networks and meeting the increasing need for more capacity, particularly in the presence of other networks operating in the same spectrum. In the context of cognitive radio research, a standard assumption has been that secondary users may search and use idle channels that are not being used by their primary users (PU). 
Although there are many existing works that focus on the algorithm design and implementation in this field, nearly all of them assume a simple independent-channel (or PU activity) model, that may not hold in practice. For instance, the operation of a low power wireless sensor network (WSN) is based on IEEE 802.15.4-radios, which uses the globally available 2.4 GHz and 868/900 MHz bands. These bands are shared by various wireless technologies (e.g. Wi-Fi, Bluetooth, RFID), as well as industrial/scientific equipment and appliances (e.g. micro-wave ovens) whose activities can affect multiple IEEE 802.15.4 channels. 
Thus, external interference can cause the channels in WSNs to be highly correlated, and the design of new algorithms and schemes in dynamic multichannel access is required to tackle this challenge.

Motivated by such practical considerations, we consider in this work a multichannel access problem with $N$ correlated channels. Each channel has two possible states: \textit{good} or \textit{bad}, and their joint distribution follow a $2^N$-states Markovian model. There is a single user (wireless node) that selects one channel at each time slot to transmit a packet.
If the selected channel is in the \textit{good} state, the transmission is successful; otherwise, there is a transmission failure. 
The goal is to obtain as many successful transmissions as possible over time. 
As the user is only able to sense his selected channel at each time slot, there is no full observation of the system available. In general, the problem can be formulated as a partially observable Markov decision process (POMDP), which is PSPACE-hard and finding the exact solution requires an exponential computation complexity~\cite{pspace-hard}. Even worse, the parameters of the joint Markovian model might not be known \emph{a-priori}, which makes it more difficult to find a good solution.

We investigate the use of Deep Reinforcement Learning, in particular, Deep Q learning, from the field of machine learning as a way to enable learning in an unknown environment as well as overcome the prohibitive computational requirements.
By integrating deep learning with Q learning, Deep Q learning or Deep Q Network (DQN)~\cite{dqn} can use a deep neural network with states as input and estimated Q values as output to efficiently learn policies for high-dimensional, large state-space problems. We implement a DQN that can find a channel access policy through online learning. 
This DQN approach is able to deal with large systems, and find a good or even optimal policy directly from historical observations without any requirement to know the system dynamics \emph{a-priori}. 
We provide a study of the optimal policy for known fixed-pattern channel switching situation and conduct various experiments showing that DQN can achieve the same optimal performance. 
We then study the performance of DQN in more complex scenarios and show through both simulations and real data trace that DQN is able to find superior, near-optimal policies. 
In addition, we also design an adaptive DQN framework that is able to adapt to 
time-varying, dynamic environments, and validate through simulations that the proposed approach can be aware of the environment change and re-learn the optimal policy for the new environment.

The rest of the paper is organized as follows.
In section~\ref{sec:related-work}, we discuss related work in the Dynamic Multichannel Access field. 
In section~\ref{sec:problem-formulation}, we formulate the dynamic multichannel access problem when channels are potentially correlated.
In section~\ref{sec:myopic-whittle}, a Myopic and a Whittle Index-based heuristic policies are presented for independent channels.
In section~\ref{sec:dqn}, we present the DQN framework to solve the problem through online learning. We present the optimal policy study on the known fixed-pattern channel switching situation in section~\ref{sec:optimal-policy-deterministic-switching}, and show through simulations that DQN can achieve the same optimal performance in section~\ref{sec:simulation-deterministic-switching}. 
We present the experiment and evaluation results of DQN in more complex situations through both simulations and real-data trace in section~\ref{sec:evaluation}. 
We propose an adaptive DQN approach in section~\ref{sec:adaptive-dqn} and conclude our work in section~\ref{sec:conclusion}.

\section{Related Work}
\label{sec:related-work}
The dynamic multichannel access problem has been widely studied. But unlike many decision making problems, such as vertical handoff in heterogeneous networks~\cite{handoff} and power allocation in energy harvesting communication systems~\cite{power}, that can be modeled as MDP,  dynamic multichannel problem is modeled as a POMDP, as channels are generally modeled as (two-state) Markov chains and a user has only partial observations of the system. And finding an optimal channel access policy has exponential time and space complexities. To overcome the prohibitive computation complexity, a Myopic policy and its performance are first studied in~\cite{myopic_1} when channels are independent and identically distributed (i.i.d.). The Myopic policy is shown to have a simple and robust round robin structure without the necessity to know the system transition probabilities except whether it is positively or negatively correlated. It is first proved in~\cite{myopic_1} that the Myopic policy is optimal when there are only two positively correlated channels in the system. Later in the subsequent work~\cite{myopic_n}, its optimality result is extended to any number of positively correlated channels and two or three negatively correlated channels. However, the Myopic policy does not have any performance guarantee when channels are correlated or follow different distributions, which is the situation considered in our work.

When channels are independent but may follow different Markov chains, the dynamic multichannel access problem can also be modeled as Restless Multi-armed bandit problem (RMAB). Each channel can be considered as an arm, and its state evolves following a Markov chain. At each time slot, a user chooses an arm with a state-dependent reward. The goal is to maximize the total expected reward over time. A Whittle Index policy is introduced in~\cite{whittle} and shares the same simple semi-universal structure and optimality result as the Myopic policy when channels are stochastically identical. Numerical results are also provided showing that the Whittle Index policy can achieve near-optimal performance when channels are nonidentical. But the Whittle Index approach cannot be applied when channels are correlated. In this work, we plan to study the multichannel access problem in the complicated correlated case.

Both the Myopic policy and the Whittle Index policy are derived under the assumption that the system transition matrix is known. When the underlying system statistics are unknown, the user must apply an online learning policy with time spent on exploration to learn the system dynamics (either explicitly or implicitly). When channels are independent, the RMAB approach can be applied and the corresponding asymptotic performance is compared with the performance achieved by a genie that has the full knowledge of the system statistics.The commonly used performance metric is called regret, which is defined as the expected reward difference between a genie and a given policy. A sublinear regret is desirable as it indicates the policy achieves the same optimal performance as the genie asymptotically. A logarithmic regret bound that grows as a logarithmic function of time $t$ is achieved in~\cite{logregret_weak_1, logregret_weak_2, logregret_weak_3} when a \emph{weak regret}\footnote{As stated in~\cite{logregret_strong_1}, ``The genie being compared with is weaker in the sense that it is aware only of the steady-state distribution for each channel, and not the full transition matrices''} is considered, and a $O(\sqrt{t})$ regret bound and a $O(\log t)$ regret bound with respect to \emph{strict regret}\footnote{As stated in~\cite{logregret_strong_1}, ``Comparing the performance of a policy to the genie that knows the probability transition matrices for each channel and can thus perform optimally''} is achieved in~\cite{rootregret_strong_1} and~\cite{logregret_strong_1} respectively. However, all these prior RMAB works are based on the independent channel assumption, which cannot be generalized for correlated channels.

In recent years, some works began to focus on the more practical and complex problem where both the system statistics is unknown and the channels are correlated. 
Q-learning, one of the most popular reinforcement learning approaches, is widely used as it is a model-free method that can learn the policy directly. 
The authors in~\cite{qlearning_seq} apply Q-learning to design channel sensing sequences, while in~\cite{qlearning_imperfect} it is shown that Q-learning can also take care of imperfect sensing.
Additionally, the work~\cite{qlearning_experiment} uses universal software radio peripheral (USRP) and GNU radio units to implement and evaluate Q-learning in a multi-hop cognitive radio network testbed. 
However, all these works assume that the system state is fully observable and formulate the problem as an MDP, which significantly reduces the state space so that  Q-learning can be easily implemented by using a look-up table to store/update Q-values. 
Since a user is only able to observe the state of the chosen channel at each time slot in our work, the current state of the system is not fully observable and our problem falls into the framework of POMDP. 
When updating Q-values, the original state space cannot be directly used because of its partial observability. 
Instead, one could consider using either the belief or a number of historical partial observations. 
This can lead to a very large state space, which makes it impossible to maintain a look-up Q table. 
New methods able to find approximations of Q-values are required to solve the large space challenge. 

In recent years, Reinforcement learning, including Q learning, has been integrated with advanced machine learning techniques, particularly deep learning, to tackle difficult high-dimensional problems~\cite{levine2015end,assael2015data,ba2014multiple}. 
In 2013, Google Deepmind used a deep neural network, called DQN, to approximate the Q values in Q learning that overcomes the limitation of the state space of the traditional look-up table approach. 
In addition, this deep neural network method also provides an end-to-end approach that an agent can learn a policy directly from his observations.
In this work, we formulate the dynamic multi-channel access problem as a POMDP and employ DQN to solve this problem. To the best of our knowledge, this is the first study and implementation of DQN in the field of dynamic multi-channel access.

\section{Problem Formulation}
\label{sec:problem-formulation}

Consider a dynamic multichannel access problem where there is a single user dynamically choosing one out of $N$ channels to transmit packets. 
Each channel can be in one of two states: \textit{good} ($1$) or \textit{bad} ($0$). 
Since channels may be correlated, the whole system can be described as a $2^N$-state Markov chain. 
At the beginning of each time slot, a user selects one channel to sense and transmit a packet. 
If the channel quality is good, the transmission succeeds and the user receives a positive reward ($+1$).
Otherwise, the transmission fails and the user receives a negative reward ($-1$). 
The objective is to design a policy that maximizes the expected long-term reward.

Let the state space of the Markov chain be $\mathcal{S} = \{\mathbf{s}_1, ..., \mathbf{s}_{2^N}\}$.
Each state $\mathbf{s}_i$ ($i \in \{1, ..., 2^N\}$) is a length-$N$ vector $[s_{i1}, ..., s_{iN}]$, where $s_{ik}$ is the binary representation of the state of channel $k$: good ($1$) or bad ($0$). 
The transition matrix of the Markov chain is denoted as $\mathbf{P}$. 
Since the user can only sense one channel and observe its state at the beginning of each time slot, the full state of the system, i.e., the states of all channels, is not observable. 
However, the user can infer the system state according to his sensing decisions and observations. 
Thus, the dynamic multichannel access problem falls into the general framework of POMDP. 
Let $\Omega(t) = [\omega_{\mathbf{s}_1}(t),..., \omega_{\mathbf{s}_{2^N}}(t)]$ represent the belief vector maintained by the user, where $\omega_{\mathbf{s}_i}(t)$ is the conditional probability that the system is in state $\mathbf{s}_i$ given all previous decisions and observations. 
Given the sensing action $a(t) \in \{1, ..., N\}$ representing which channel to sense at the beginning of time slot $t$, the user can observe the state of channel $a(t)$, denoted as $o(t) \in \{0, 1\}$. 
Then, based on this observation, he can update the belief vector at time slot $t$, denoted as $\hat{\Omega}(t) = [\hat{\omega}_{\mathbf{s}_1}(t),..., \hat{\omega}_{\mathbf{s}_{2^N}}(t)]$.
The belief of each possible state $\hat{\omega}_{\mathbf{s}_i}(t)$ is updated as follows:
\begin{equation}
 \hat{\omega}_{\mathbf{s}_i}(t) =  \begin{cases} 
      \frac{\omega_{\mathbf{s_i}}(t)\mathbbm{1}(\mathbf{s}_{ik}(t)=1)}{\sum_{i=1}^{2^N} \omega_{\mathbf{s}_i}(t)\mathbbm{1}(\mathbf{s}_{ik}(t)=1)} & a(t) = k, o(t) = 1 \\
     \frac{\omega_{\mathbf{s}_i}(t)\mathbbm{1}(\mathbf{s}_{ik}(t)=0)}{\sum_{i=1}^{2^N} \omega_{\mathbf{s}_i}(t)\mathbbm{1}(\mathbf{s}_{ik}(t)=0)} & a(t) = k, o(t) = 0  
   \end{cases}
\end{equation}
where $\mathbbm{1}(.)$ is the indicator function.

Combining the newly updated belief vector $\hat{\Omega}(t)$ for time slot $t$ with the system transition matrix $\mathbf{P}$, the belief vector for time slot $t+1$ can be updated as:
\begin{equation}
\label{eqn:belief}
\Omega(t+1) = \hat{\Omega}(t) \mathbf{P}
\end{equation}

A sensing policy $\pi : \Omega(t) \rightarrow a(t)$ is a function that maps the belief vector $\Omega(t)$ to a sensing action $a(t)$ at each time slot $t$. 
Given a policy $\pi$, the long-term reward considered in this paper is the expected accumulated discounted reward over infinite time horizon, defined as:
\begin{equation}
\mathbb{E}_{\pi} [\sum_{t=1}^{\infty} \gamma^{t-1} R_{\pi(\Omega(t))}(t) | \Omega(1)]
\end{equation}
where $0 \leq \gamma < 1$ is a discounted factor, $\pi(\Omega(t))$ is the action (i.e., which channel to sense) at time $t$ when the current belief vector is $\Omega(t)$, and $R_{\pi(\Omega(t))}(t)$ is the corresponding reward.

If no information about the initial distribution of the system state is available, one can assume the initial belief vector $\Omega(1)$ to be the stationary distribution of the system. 
Our objective is to find a sensing policy $\pi^*$ that maximizes the expected accumulated discounted reward over infinite time
\begin{equation} 
\pi^* = \argmax_{\pi} \mathbb{E}_{\pi} [\sum_{t=1}^{\infty} \gamma^{t-1} R_{\pi(\Omega(t))}(t) | \Omega(1)]
\end{equation}

As the dynamic multichannel access problem is a POMDP, the optimal sensing policy $\pi^*$ can be found by considering its belief space and solving an augmented MDP instead. Let $\mathcal{B}$ represent the belief space, and let $V^*(b)$ be the maximum expected accumulated discounted reward from the optimal policy $\pi^*$ with initial belief as $b$. Then for all belief $b\in \mathcal{B}$, we have the following Bellman optimality equation
\begin{equation}
\begin{aligned}
V^*(b) = \max_{k=1, ..., N} \Bigg{\{}\!\!\sum_{i=1}^{2^N}\omega_{\mathbf{s}_i} \mathbbm{1}(\mathbf{s}_{ik}=1) &+ \gamma  \sum_{i=1}^{2^N}\omega_{\mathbf{s}_i} \mathbbm{1}(\mathbf{s}_{ik}=1) V^*(T(b|a=k, o=1)) \\
&+  \gamma  \sum_{i=1}^{2^N}\omega_{\mathbf{s}_i} \mathbbm{1}(\mathbf{s}_{ik}=0) V^*(T(b|a=k, o=0))\Bigg{\}}
\end{aligned}
\end{equation}
 where the $T(b|a, o)$ is the updated belief at given the action $a$ and observation $o$ as in Eq.~(\ref{eqn:belief}).
 
In theory, the value function $V^*(b)$ together with the optimal policy $\pi^*$ can be found via value iteration approach. However, since there are multiple channels and they might be correlated, the belief space becomes a high-dimensional space. For instance, in a typical multichannel WSN based on the widely used IEEE 802.15.4-2015 standard~\cite{ieee802154-2015}, nodes have to choose one out of $16$ available channels to sense at each time slot. If we consider the potential correlations among channels and simplify each channel's condition to be in only two states: good or bad, the state space size becomes $2^{16}$. As the belief represents a probability distribution function over all possible states, it also becomes high dimensional, which increases computation cost.

\begin{figure}
\vspace{-0.8cm}
\centering
\begin{minipage}{.45\textwidth}
  \centering
  \includegraphics[width=1\linewidth]{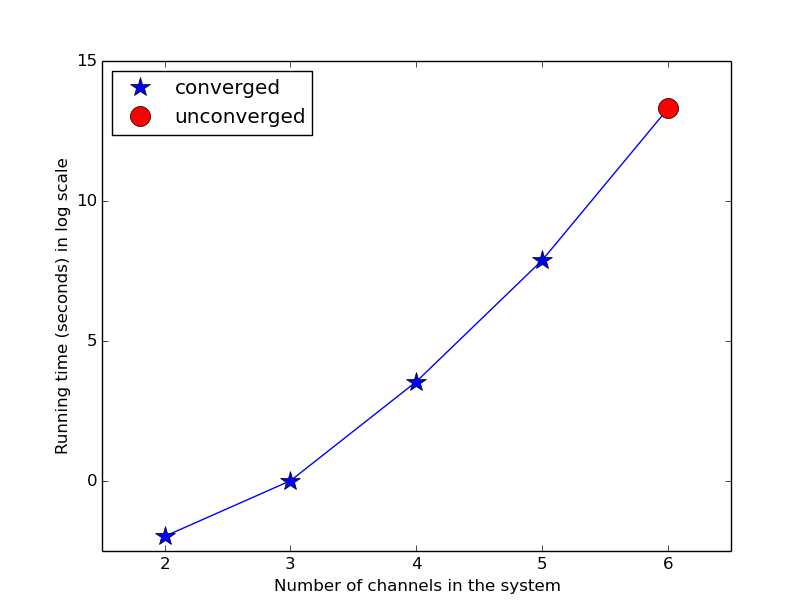}
  \captionof{figure}{Running time (seconds) in log scale of the POMDP solver as we vary the number of channels in the system}
  \label{fig:run_time}
\end{minipage}%
\hfill
\begin{minipage}{.45\textwidth}
  \centering
  \includegraphics[width=0.8\linewidth]{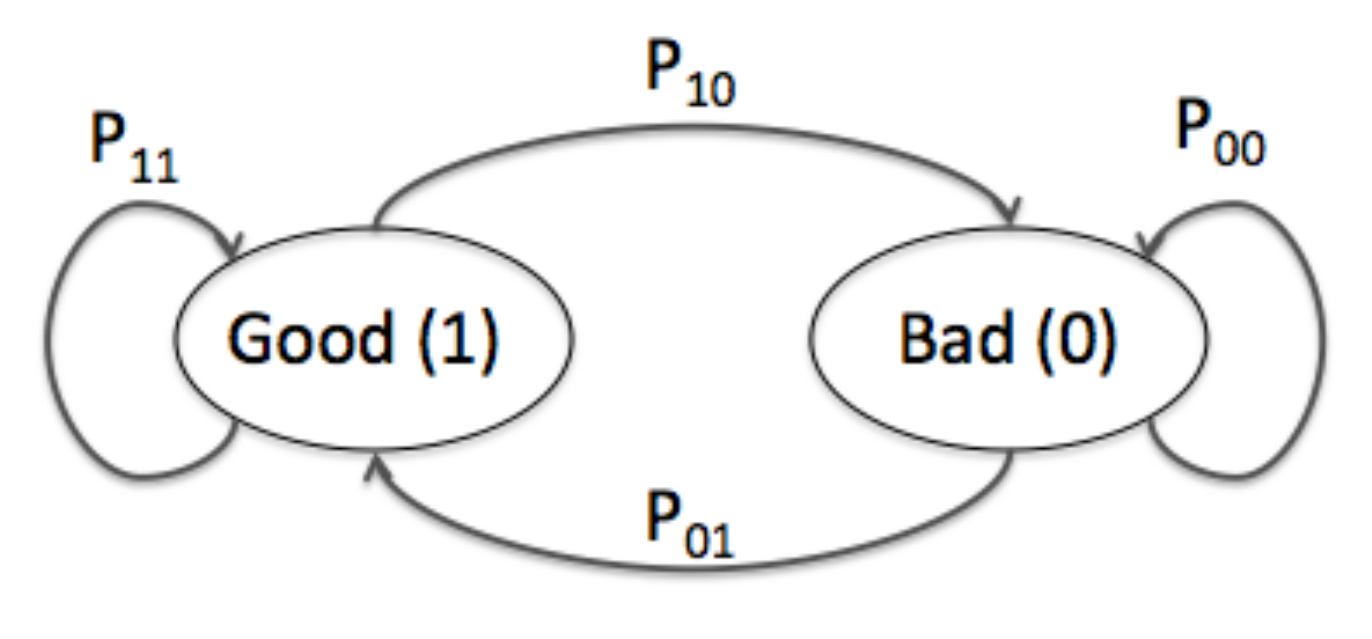}
  \captionof{figure}{Gilbert-Elliot channel model}
  \label{fig:chann_fig}
\end{minipage}
\vspace{-0.5cm}
\end{figure}

Even worse, the infinite size of the continuous belief space and the impact of the current action on the future reward makes POMDP PSPACE-hard, which is even less likely to be solved in polynomial time than NP-hard problems~\cite{pspace-hard}. To exemplify the time complexity of solving such POMDP problem, we simulate the multichannel access problem with known system dynamics and use a POMDP solver called SolvePOMDP~\cite{pomdp} to find its optimal solution. In Figure~\ref{fig:run_time}, we show the run-time as we increase the number of channels in the system.  When the number of channels is higher than $5$, the POMDP solver can not converge after a long interval, and it gets terminated when the run-time exceeds the time limit. 

All these factors make it impossible to find the optimal solution to a POMDP in general, and many existing works~\cite{myopic_1, myopic_n, whittle, logregret_weak_1, logregret_weak_2, logregret_weak_3, logregret_strong_1, rootregret_strong_1} attempt to address this challenge of prohibitive computation by considering either simpler models or approximation algorithms.

\section{Myopic Policy and Whittle Index}
\label{sec:myopic-whittle}

In the domain of dynamic multichannel access, there are many existing works on finding the optimal/near-optimal policy with low computation cost when the channels are independent and system statistics ($\mathbf{P}$) is known. The Myopic policy and the Whittle Index policy are two effective and easy-to-implement approaches for this settings.

\subsection{Myopic Policy}

A Myopic policy only focuses on the immediate reward obtained from an action and ignores its effects in the future.
Thus the user always tries to select a channel which gives the maximized expected immediate reward
\begin{equation}
\hat{a}(t) = \argmax_{k=1, ..., N} \sum_{i=1}^{2^N} \omega_{\mathbf{s}_i}(t) \mathbbm{1}(\mathbf{s}_{ik}(t)=1)
\end{equation}

The Myopic policy is not optimal in general. 
Researchers in~\cite{myopic_1},~\cite{myopic_n} have studied its optimality when $N$ channels are independent and statistically identical Gilbert-Elliot channels that follow the same $2$-state Markov chain with the transition matrix as $\icol{p_{00} \quad p_{01}\\p_{10} \quad p_{11}}$, as illustrated in Fig.~\ref{fig:chann_fig}.
It is shown that the Myopic policy is optimal for any number of channels when the channel state transitions are positively correlated, i.e., $p_{11} \geq p_{01}$. 
The same optimal result still holds for two or three channels when channel state transitions are negatively correlated,  i.e., $p_{11} < p_{01}$.
In addition, the Myopic policy has a simple robust structure that follows a round-robin channel selection procedure. 


\subsection{Whittle Index Based Heuristic Policy}

When channels are independent, the dynamic multichannel access problem can also be considered as a restless multi-armed bandit problem (RMAB) if each channel is treated as an arm. 
An index policy assigns a value to each arm based on its current state and chooses the arm with the highest index at each time slot. Similarly, the index policy does not have optimality guarantee in general.

In~\cite{whittle}, the Whittle Index is introduced in the case when $\mathbf{P}$ is known and all channels are independent but may follow different $2$-state Markov chain models. 
In this case, the Whittle Index policy can be represented as a closed-form solution, and it has the same optimal result as the Myopic policy: the Whittle Index policy is optimal for any number of channels when channels are identical and positively correlated, or for two or three channels when channels are negatively correlated. 
In addition, when channels follow identical distributions, the Whittle Index policy has the same round-robin structure as the Myopic policy. 

When channels are correlated, the Whittle Index cannot be defined and thus the Whittle Index policy cannot be directly applied to our problem. 
To leverage its simplicity, we propose an heuristic that ignores the correlations among channels and uses the joint transition matrix $\mathbf{P}$ and Bayes' Rule to compute the $2$-state Markov chain for each individual channel. Assume that for channel $k$, the transition matrix is represented as $p(c_k^{t+1}=m|c_k^{t}=n)$, where $m,n \in \{0,1\}$ (bad or good). 
Then, based on Bayes' Rule we have,
\begin{equation}
p(c_k^{t+1}=m|c_k^{t}=n) = \frac{p(c_k^{t+1}=m, c_k^{t}=n)}{p(c_k^{t}=n)} = 
\frac{\sum_{j=1}^{2^N}\sum_{i=1}^{2^N}p(\mathbf{s}_j|\mathbf{s}_i)p(\mathbf{s}_i)\mathbbm{1}(s_{jk}=m)\mathbbm{1}(s_{ik}=n)}{\sum_{i=1}^{2^N}p(\mathbf{s}_i) \mathbbm{1}(s_{ik}=n)}
\end{equation}
where $p(\mathbf{s}_i)$ is the stationary distribution and $p(\mathbf{s}_j|\mathbf{s}_i)$ is the transition probability from state $\mathbf{s}_i$ to state $\mathbf{s}_j$ defined in $\mathbf{P}$. After each channel model is found, we can apply the Whittle Index policy.

The Myopic policy and the Whittle Index policy are easy to implement in practice, as both of them have polynomial run-time. And in the case of independent channels, the Myopic and the Whittle Index policies can achieve optimality under certain conditions. 
However, so far to the best of our knowledge there are no easy-to-implement policies applicable to the general case where channels are correlated.
Moreover, both policies require the prior knowledge of the system's transition matrix, which is hard to obtain beforehand in practice. 
Thus, we need to come up with a new approach that copes with these challenges.

\section{Deep Reinforcement Learning Framework}
\label{sec:dqn}
When channels are correlated and system dynamics are unknown, there are two main approaches to tackle the dynamic multichannel access problem: (i) Model-based approach: first estimating the system model from observations and then either solve it by following the dynamic programming method in Section III or apply some computationally efficient heuristic algorithm such as the Myopic policy and the Whittle Index policy (which have polynomial run-time); (ii) Model-free approach: learn the policy directly through interactions with the system without estimate the system model. The model-based approach is less favored since the user can only observe one channel at a time slot and the limited observation capability may result in a bad system model estimation. Even worse, even if the system dynamics is well estimated, solving a POMDP in a large state space is always a bottleneck as the dynamic programming method has exponential time complexity (as explained in Section III) and the heuristic approaches do not have any performance guarantee in general. All these challenges motivate us to follow the model-free approach, which, by incorporating the idea of Reinforcement Learning, can learn directly from observations without the necessity of finding an estimated system model and can be easily extended to very large and complicated systems.
 
\subsection{Q-Learning}
We focus on Reinforcement Learning paradigm, Q-learning~\cite{Q-learning} specifically, to incorporate learning in the solution for the dynamic multichannel access problem. The goal of Q-learning is to find an optimal policy, i.e., a sequence of actions that maximizes the long-term expected accumulated discounted reward. 
Q-learning is a value iteration approach and the essence is to find the Q-value of each state and action pairs, where the state $\mathbf{x}$ is a function of observations (and rewards) and the action $a$ is some action that a user can take given the state $\mathbf{x}$. 
The Q-value of a state-action pair $(\mathbf{x}, a)$ from policy $\pi$, denoted as $Q^{\pi}(\mathbf{x}, a)$, is defined as the sum of the discounted reward received when taking action $a$ in the initial state $\mathbf{x}$ and then following the policy $\pi$ thereafter. $Q^{\pi^*} (\mathbf{x}, a)$ is the Q-value with initial state $\mathbf{x}$ and initial action $a$, and then following the optimal policy $\pi^*$. Thus, the optimal policy $\pi^* $ can be derived as 
\begin{equation}
\pi^*(\mathbf{x}) = \argmax_{a} Q^{\pi^*}(\mathbf{x}, a), \forall \mathbf{x}
\end{equation}

One can use online learning method to find $Q^{\pi^*}(\mathbf{x}, a)$ without any knowledge of the system dynamics.  Assume at the beginning of each time slot, the agent takes an action $a_t \in \{1, ..., N\}$ that maximizes its Q-value of state-action pair $(\mathbf{x}_t, a_t)$ given the state is $\mathbf{x}_t$, and gains a reward $r_{t+1}$.
Then the online update rule of Q-values with learning rate $0<\alpha<1$ is given as follows:
\begin{equation}
\label{eqn:q-update}
Q(\mathbf{x}_t, a_t) \leftarrow  Q(\mathbf{x}_t, a_t)  + \alpha [r_{t+1} + \gamma \max_{a_{t+1}}Q(\mathbf{x}_{t+1}, a_{t+1}) 
- Q(\mathbf{x}_t, a_t)]
\end{equation}

It has been shown that in the MDP case, if each action is executed in each state an infinite number of times on an infinite run and the learning rate $\alpha$ decays appropriately, the Q-value of each state and action pair will converge with probability 1 to the optimal $Q^{\pi^*}$, and thus the optimal policy can be found~\cite{qlearning}. 

In the context of the dynamic multichannel access, the problem can be converted to an MDP when considering the belief space, and Q-learning can be applied consequently. However, this approach is impractical since the belief update is maintained by knowing the system transition matrix $\mathbf{P}$ \textit{a-priori}, which is hardly available in practice. 
Instead, we apply Q-learning by directly considering the history of observations and actions. We define the state for the Q-learning at time slot $t$ as a combination of historical selected channels as well as their observed channel conditions over previous $M$ time slots, i.e., $\mathbf{x}_t = [a_{t-1}, o_{t-1}, ..., a_{t-M}, o_{t-M}]$.
Then we can execute the online learning following Eq.~(\ref{eqn:q-update}) to find the sensing policy. Intuitively, the more historical information we consider (i.e., the larger $M$ is), the better Q-learning can learn.

\subsection{Deep Q-Network}

Q-learning works well when the problem's state-action spaces are small, as a look-up table can be used to execute the update rule in Eq.~(\ref{eqn:q-update}). 
But this is impossible when the state-action space becomes very large. Even worse, since many states are rarely visited, their corresponding Q-values are seldom updated. This causes Q learning takes a very long time to converge. 

In this work, the state space size of Q-learning is $(2N)^M$, which grows exponentially with $M$. This is because the state of Q-learning is defined as a combination of observations and actions over past $M$ time slots. In a single time slot, the number of possible observations is $2N$, as the user can only sense one out of $N$ channels and each channel has $2$ possible states. We do not consider the size of action space as action information is implicitly included in the observation. Thus, the state size of Q-learning is the number of all possible combinations of observations over previous $M$ time slots, which is $(2N)^M$. As we mentioned before, the number of previous time slots $M$ is also required to be large so that Q-learning can capture enough system information and learn better. This can cause the state space of Q-learning become very large, which prohibits using a traditional look-up table approach.

Researchers have proposed both linear and non-linear Q-value approximations to overcome the state space size limit.
In 2013, DeepMind developed a Deep Q-Network (DQN), which makes use of a deep neural network to approximate the Q-values, and it achieves human-level control in the challenging domain of classic Atari 2600 games~\cite{dqn}. 

A neural network is a biologically-inspired programming paradigm organized in layers. Each layer is made up of a number of nodes known as neurons, each of which executes an `activation function'. Each neuron takes the weighted linear combination of the outputs from neurons in the previous layer as input and outputs the result from its nonlinear activation function to the next layer. The networked-neuron architecture enables the neural network to be capable of approximating nonlinear functions of the observational data.  A deep neural network is a neural network that can be considered as a deep graph with many processing layers. A deep neural network is able to learn from low-level observed multi-dimensional data and find its success in areas such as computer vision and natural language processing~\cite{dnn_cv, dnn_nlp}.


DQN combines Q-learning with deep learning, and the Q-function is approximated by a deep neural network called Q-network that takes the state-action pair as input and outputs the corresponding Q-value. Q-network updates its weights $\mathbf{\theta}$ at each iteration $i$ to minimize the loss function $L_i(\mathbf{\theta}_i) = \mathbb{E} [(y_i - Q(\mathbf{x},a;\mathbf{\theta}_i))^2]$, where $y_i= \mathbb{E}[r + \gamma \max_{a'} Q(\mathbf{x}', a';\mathbf{\theta}_{i-1})]$ is derived from the same Q-network with old weights $\mathbf{\theta}_{i-1}$ and new state $\mathbf{x}'$ after taking action $a$ from state $\mathbf{x}$. 

Since we directly use the previous historical observations and actions as the state for the Q-learning, the state space becomes exponentially large as we increase the considered historical information, and a traditional look-up table approach to maintain Q values does not work well. Therefore, a DQN implementation is needed to help to find a tractable policy implementation in the dynamic multichannel access problem.

\section{Optimal Policy for Known Fixed-Pattern Channel Switching}
\label{sec:optimal-policy-deterministic-switching}

To study the performance of DQN, we first consider a situation when all the $N$ channels in the system can be divided into several independent subsets and these subsets take turns to be activated following a fixed pattern. 
We assume at each time slot, only a single subset is activated such that all channels in the activated subset are good and all channels in inactivated subsets are bad. At each time slot, with probability $p$ ($0 \leq p \leq 1$) the next following subset is activated, and with probability $1-p$ the current subset remains activated. We assume the activation order of the subsets is fixed and will not change over time.

In this section, we assume that the subset activation order, the activation switching probability $p$ as well as the initially activated subset are known \emph{a-priori}. 
The optimal policy can be found analytically and is summarized in Theorem 1. This serves as a baseline to evaluate the performance of DQN implementation in the next section.

\begin{theorem}
When the system follows a fixed-pattern channel switching, if the activation order, switching probability $p$ and the initial activation subset are known, the optimal channel access policy follows  Algorithm~\ref{alg:optPolicy_1} or Algorithm~\ref{alg:optPolicy_2} depending on the value of $p$.
\end{theorem}

\vspace{-0.6cm}
\begin{minipage}[t]{.45\textwidth}
\begin{algorithm}[H]
    \small
    \caption{\small{Optimal Policy when $0.5 \leq p \leq 1$}}
    \label{alg:optPolicy_1}
    \begin{algorithmic}[1]
    \State At the beginning of time slot $0$, choose a channel in the initial activated subset $C_1$
    \For{$n=1,2,\ldots$ }
        \State At the beginning of time slot $n$, 
        \If{The previous chosen channel is good}
            \State Choose a channel in the next activated subset according to the subset activation order
        \Else
            \State Stay in the same channel
        \EndIf
    \EndFor
    \end{algorithmic}
\end{algorithm}
\end{minipage}%
\hfill
\begin{minipage}[t]{.45\textwidth}
 \begin{algorithm}[H]
    \small
    \caption{\small{Optimal Policy when $0 \leq p < 0.5$}}
    \label{alg:optPolicy_2}
    \begin{algorithmic}[1]
 \State At the beginning of time slot $0$, choose a channel in the initial activated subset $C_1$
 \For{$n=1,2,\ldots$ }
 \State At the beginning of time slot $n$. 
 \If{The previous chosen channel is good}
 \State Stay in the same channel
 \Else
 \State Choose a channel in the next activated subset according to the subset activation order
 \EndIf
 \EndFor
 \end{algorithmic}
 \end{algorithm}
\end{minipage}



\vspace{1cm}
\begin{IEEEproof}
Assume the currently activated subset at each time slot is known \emph{a-priori}. Then the problem can be modeled as a fully-observable MDP, and the corresponding optimal policy can be found by finding and comparing Q values of all possible state-action pairs.

Assume all $N$ channels in the system form $M$ independent subsets, thus there are $M$ states in total. The subsets are indexed according to their fixed activation order as $C_1, C_2, \ldots, C_M$, where $C_1$ is the initial activation subset at the start of the system. Note the channel subset activation order is circular so that $C_M$ is followed by $C_1$ in the order. The corresponding system state at a time slot is represented as $S_i$ ($1 \leq i \leq M$) when channel subset $C_i$ is activated. Let $p(S_j|S_i)$ be the transition probability from state $S_i$ to state $S_j$ ($i,j \in \{1, \ldots, M\}$) of the Markov chain, and we have:
\begin{equation}
        p(S_j|S_i)=
        \begin{cases}
            p, & j=i+1 \\
            1-p, & j=i 
        \end{cases}
        \label{eq-markov-chain}
\end{equation}

Then the corresponding Q-value of the optimal policy starting with state $S_i$ and action $a$ is:
\begin{equation}
Q^*(S_i, a) = \sum_{j=1}^{M} p(S_j|S_i)[R(S_i,a) + \gamma V^*(S_j)]
\label{eq-q-value-optimal-policy}
\end{equation}
where $R(S_i,a)$ is the immediate reward, i.e. either $+1$ if the chosen channel is good or $-1$ if the chosen channel is bad. $V^*(S_j)$, defined as $\max_a Q^*(S_i,a)$, represents the expected accumulated discounted reward given by an optimal policy over infinite time horizon with initial state as $S_j$.

Taking Eq.~(\ref{eq-markov-chain}) into Eq.~(\ref{eq-q-value-optimal-policy}), we have
\begin{equation}
\label{eqn:q_value}
\begin{aligned}
Q^*(S_i, a) = 
&\begin{cases}
p\cdot 1+(1-p)\cdot(-1)+c, & a \in C_{i+1}\\
p\cdot(-1)+(1-p)\cdot 1+c, & a \in C_{i}\\
-1 + c, & \text{otherwise}
\end{cases}
=&\begin{cases}
2p-1+c, & a \in C_{i+1}\\
1-2p+c, & a \in C_{i}\\
-1 + c, & \text{otherwise}
\end{cases}
\end{aligned}\\
\end{equation}
where $c=\gamma[pV^*(S_{i+1})+(1-p)V^*(S_{i})]$, which does not depend on the action.

Since the optimal action $a^*(S_i)$  for each state $S_i$ is $a^*(S_i) = \argmax_a Q^*(S_j,a)$, the optimal action to maximize the Q value of a given state $S_i$ in Eq.~(\ref{eqn:q_value}) is 
\begin{equation}
\label{eqn:optimal_policy}
a^*(S_i) = 
\begin{cases}
\text{any channel in $C_{i+1}$}, & 0.5 \leq p \leq 1 \\
\text{any channel in $C_{i}$}, & 0 \leq p < 0.5
\end{cases}
\end{equation}

All the above analysis holds based on the assumption that the current state of each time slot is observable. As the initially activated channel subset is known, the user can initially choose a channel in this activated subset and then follow Eq.~(\ref{eqn:optimal_policy}) afterward. Based on the observation of the chosen channel the user is guaranteed to know what the current state is: if the chosen channel is good, the currently activated subset is the subset containing the chosen channel; otherwise, the currently activated subset is the subset prior to the chosen channel's subset in the activation order. Thus, the current state of the MDP is fully observable, and the optimality of the policies in Alg.~\ref{alg:optPolicy_1} and Alg.~\ref{alg:optPolicy_2} derived from Eq.~(\ref{eqn:optimal_policy}) is achieved. 
\end{IEEEproof}

It turns out that the optimal policy for the fixed-pattern channel switching shares a similarly simple and robust structure with the Myopic policy in~\cite{myopic_1}: the optimal policy has a round-robin structure (in terms of the channel subset activation order) and does not require to know the exact value of $p$ except whether it is above/below $0.5$. This semi-universal property makes the optimal policy easy to implement in practice and robust to mismatches of system dynamics.

\section{Experiment and Evaluation of Learning for Unknown Fixed-Pattern Channel Switching}
\label{sec:simulation-deterministic-switching}

Having derived the optimal policy for fixed-pattern channel switching when one has a full knowledge of the system statistics in the previous section, we implement a DQN in this section and study how it performs in the fixed-pattern channel switching even without any prior knowledge of the system statistics. We first present details of our DQN implementation and then evaluate its performance through three experiments. 

\subsection{DQN Architecture}

We design a DQN by following the \emph{Deep Q-learning with Experience Replay Algorithm}~\cite{dqn} and implement it in TensorFlow~\cite{tensorflow}.
The structure of our DQN is finalized as a fully connected neural network with each of the two hidden layers containing $200$ neurons\footnote{Generally speaking, deciding the number of hidden layers and the number of neurons in a layer needs many trials and errors. But we follow some general guidance provided in~\cite{Heaton}. We choose a two-hidden-layers neural network as it ``can represent an arbitrary decision boundary to arbitrary accuracy with rational activation functions and can approximate any smooth mapping to any accuracy." And to decide the number of neurons in each layer, one of the rules of thumb methods is that ``The number of hidden neurons should be between the size of the input layer and the size of the output layer." We have tried a different number of neurons between $16$ (output layer size) and $256$ (input layer size), and the network structure with $200$ neurons provided a good performance with small training time.}. The activation function of each neuron is Rectified Linear Unit (\textit{ReLU}), which computes the function $f(x)=\max(x,0)$. The state of the DQN is defined as the combination of previous actions and observations over previous $M$ time slots, which serves as the input to the DQN. And the considered number of historical time slots is the same as the number of channels in the system, i.e., $M=N$. A vector of length $N$ is used to represent the observation at a time slot, where each item in the vector indicates the quality of the corresponding channel. If channel $i$ is selected, the value of the $i$th item in the vector is $1$ if the channel quality is good or $-1$ if the channel quality is bad; otherwise, we use $0$ to indicate that channel $i$ is not selected. And this vector implicitly contains action information, as a non-zero item in the vector indicates the corresponding channel is selected. The output of the DQN is a vector of length $N$, where the $i$th item represents the Q value of a given state if channel $i$ is selected. We apply the $\epsilon$-greedy policy with $\epsilon$ fixed as $0.1$ to balance the exploration and exploitation, i.e., with probability $0.1$ the agent selects uniformly a random action, and with probability $0.9$ the agent chooses the action that maximizes the Q value of a given state. A technique called $\emph{Experience Replay}$ is introduced in~\cite{dqn} to break correlations among data samples and make the training stable and convergent. 
At each time slot $t$ during training, when an action $a_t$ is taken given the state is $\mathbf{x}_t$, the user gains a corresponding reward $r_t$ and the state is updated to $\mathbf{x}_{t+1}$, a piece of record $(\mathbf{x}_t, a_t, r_t, \mathbf{x}_{t+1})$ is stored into a place called replay memory. When updating the weights $\mathbf{\theta}$ of the DQN, a minibatch of $32$ samples are randomly selected from the replay memory to compute the loss function, and then a recently proposed Adam algorithm~\cite{adam} is used to conduct the stochastic gradient descent to update the weights (details on the hyperparameters are listed in Table~\ref{tab:hyperparameters}). In the following experiment settings, we consider a system of $16$ channels, i.e., $N=16$, which is a typical multichannel WSN.

\begin{table}
\centering
\caption{List of DQN Hyperparameters}
\label{tab:hyperparameters}
 \begin{tabular}{cp{0.25\textwidth}}\hline
Hyperparameters & \qquad \qquad Values \\
\hline
$\epsilon$ & \qquad \qquad $0.1$\\
Minibatch size & \qquad \qquad $32$\\
Optimizer & \qquad \qquad Adam \\
Activation Function & \qquad \qquad ReLU \\
Learning rate & \qquad \qquad $10^{-4}$\\
Experience replay size & \qquad \qquad $1,000,000$\\
$\gamma$ & \qquad \qquad $0.9$\\
\hline
 \end{tabular}
\end{table}

\subsection{Single Good Channel, Round Robin Switching Situation}
\begin{figure}
\vspace{-1cm}
\centering
\begin{minipage}{.45\textwidth}
  \centering
  \includegraphics[width=.4\linewidth]{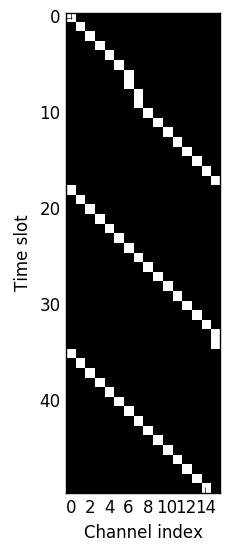}
  \captionof{figure}{A capture of a single good channel, round robin switching situation over $50$ time slots}
  \label{fig:illu_round}
\end{minipage}%
\hfill
\begin{minipage}{.45\textwidth}
\vspace{1.2cm}
  \centering
  \includegraphics[width=1\linewidth]{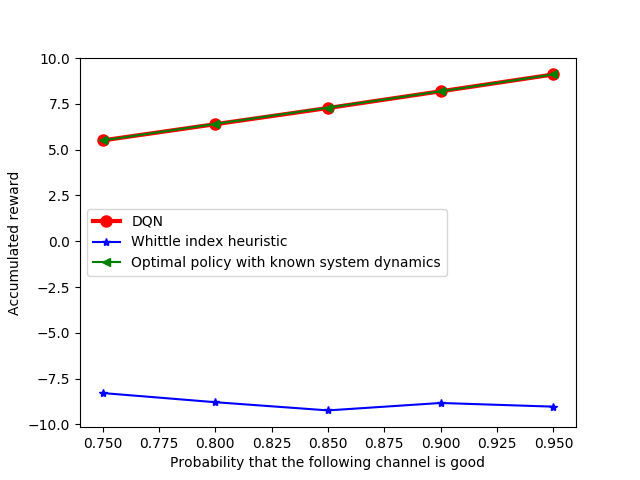}
  \captionof{figure}{Average discounted reward as we vary the switching probability $p$ in the single good channel, round robin switching}
  \label{fig:round_robin}
\end{minipage}
\vspace{-0.5cm}
\end{figure}

We first consider a situation where there is only one good channel in the system at any time slot. The channels take turns to become good with some probability in a sequential round-robin fashion. In other words, if at time slot $t$, channel $k$ is good and all other channels are bad, then in the following time slot $t+1$, with probability $p$ the following channel $k+1$ becomes good and all others bad, and with probability $1-p$ channel $k$ remains good and all others bad. In this situation, the inherited dependence and correlation between channels are high. Actually, this is the fixed-pattern channel switching with each independent subset contains one single channel and is activated in a sequential order. In Fig.~\ref{fig:illu_round}, we provide a pixel illustration to visualize how the states of channels change in the $16$-channel system that follows a single good channel, round-robin situation over $50$ time slots. The x-axis is the index of each channel, and the y-axis is the time slot number. A white cell indicates that the corresponding channel is good, and a black cell indicates that the corresponding channel is bad.

We compare the DQN with two other policies: the Whittle Index heuristic policy and the optimal policy with known system dynamics from section~\ref{sec:optimal-policy-deterministic-switching}. The optimal policy has full knowledge of the system dynamics and serves as a performance upper bound. In the Whittle Index heuristic, the user assumes all channels are independent. For each channel, the user observes it for $10,000$ time slots and uses Maximum Likelihood Estimation (MLE) to estimate the corresponding $2$-state Markov chain transition matrix. Once the system model is estimated, Whittle Index can be applied. As can be seen in Fig.~\ref{fig:round_robin}, as the switching probability $p$ varies, DQN remains robust and achieves the same optimal performance in all five cases as the optimal policy and performs significantly better than the Whittle Index heuristic. This lies in the fact that DQN can implicitly learn the system dynamics including the correlation among channels, and finds the optimal policy accordingly. On the contrary, the Whittle Index heuristic simply assumes the channels are independent and is not able to find or make use of the correlation among channels. Moreover, as the switching probability $p$ increases, the accumulated reward from DQN also increases because there is more certainty in the system that leads to an increase in the optimal reward.

\subsection{Single Good Channel, Arbitrary Switching Situation}
\begin{figure}
\vspace{-1cm}
\centering
\begin{minipage}{.45\textwidth}
  \centering
  \includegraphics[width=.4\linewidth]{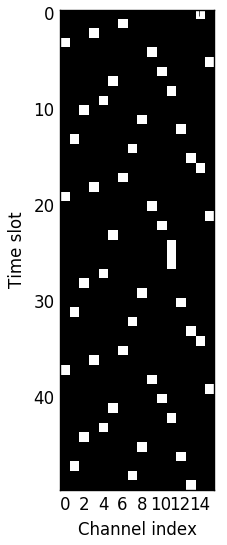}
  \captionof{figure}{A capture of a single good channel, arbitrary switching situation over $50$ time slots}
  \label{fig:illu_random}
\end{minipage}%
\hfill
\begin{minipage}{.45\textwidth}
\vspace{1.2cm}
  \centering
  \includegraphics[width=1\linewidth]{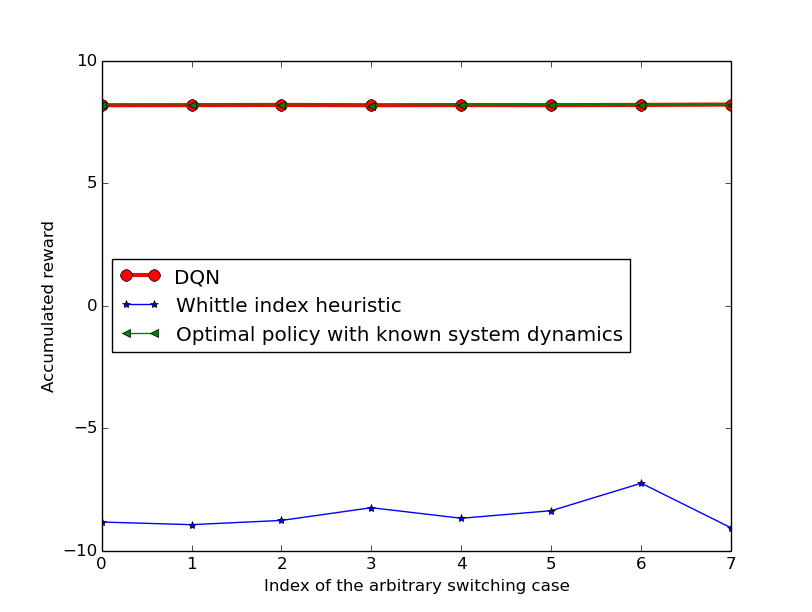}
  \captionof{figure}{Average discounted reward as we vary the switching order in the single good channel, arbitrary switching}
  \label{fig:rand}
\end{minipage}
\vspace{-0.5cm}
\end{figure}

Next, we study a situation in which there is still only one channel being good in any time slot. However, unlike the previous situation, the channels become good in an arbitrary order. 
Fig.~\ref{fig:illu_random} shows a pixel illustration of the $16$-channel system in this situation.

In the experiment, the channel-switching probability $p$ is fixed as $0.9$, and we randomly choose $8$ different arbitrary channel switching orders. As can be seen from Fig.~\ref{fig:rand}, DQN achieves the optimal performance and significantly outperforms Whittle Index heuristic in all cases. 




\subsection{Multiple Good Channels Situation}
\begin{figure}
\vspace{-1cm}
\centering
\begin{minipage}{.45\textwidth}
  \centering
  \includegraphics[width=0.5\linewidth]{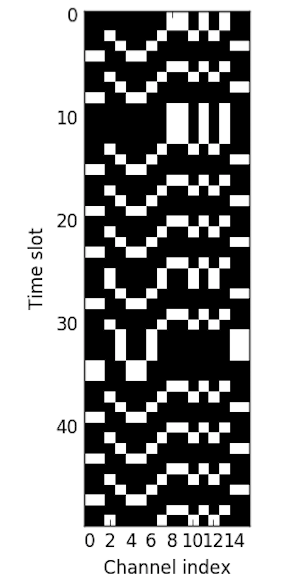}
  \captionof{figure}{A capture of a multiple good channels situation over $50$ time slots}
  \label{fig:illu_mul}
\end{minipage}%
\hfill
\begin{minipage}{.45\textwidth}
  \centering
  \includegraphics[width=1\linewidth]{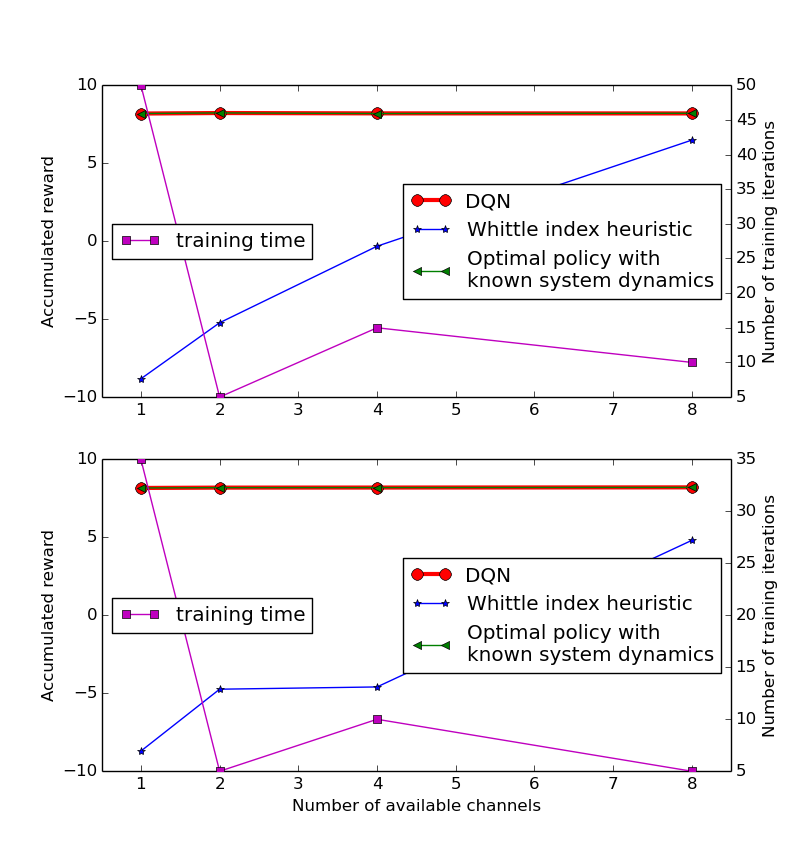}
  \captionof{figure}{Average discounted reward as we increase the number of good channels in the multiple good channels situation}
  \label{fig:mul_chann}
\end{minipage}
\vspace{-0.5cm}
\end{figure}

In this section, we investigate the situation when there may be more than one good channels in a time slot. The $16$ channels are evenly divided into several subsets, where each subset contains the same number of channels. At any time slot, there is only one subset activated where all channels in this subset are good, and channels in other inactivated subsets are bad. The subsets take turns to become available with a switching probability fixed at $0.9$. And this is the fixed-pattern channel switching with each independent subset contains one or more channels. Fig.~\ref{fig:illu_mul} shows a pixel illustration of the $16$-channel system in a multiple good channels situation.

We vary the number of channels in a subset as $1$, $2$, $4$ and $8$ in the experiment, and present the experimental result in Fig.~\ref{fig:mul_chann}. The $16$ channels in the system are in order and the subsets are activated in a sequential round-robin order in the upper graph in Fig.~\ref{fig:mul_chann}, while the channels are arranged arbitrarily and the activation order of subsets is also arbitrary in the bottom graph in Fig.~\ref{fig:mul_chann}. As can be seen, DQN always achieve the optimal performance, and the training time decreases as the number of good channels increases. This is because there is more chance to find a good channel when more good channels are available at a time slot, and the learning process becomes easier so that the DQN agent can take less time exploring and is able to find the optimal policy more quickly. This also explains why Whittle Index heuristic performs better when there are more good channels available. However, DQN significantly outperforms Whittle Index heuristic in all cases.


\section{Experiment and Evaluation of DQN for More Complex Situations}
\label{sec:evaluation}

From the results in Section~\ref{sec:simulation-deterministic-switching}, we can see that DQN outperforms Whittle Index heuristic and achieves optimal performance in the unknown fixed-pattern channel switching. Another question to ask is: can DQN achieve a good or even optimal performance in more complex and realistic situations? To answer this question and at the same time provide a better and deeper understanding of DQN, we have re-tuned our neural network structure to become a fully connected neural network with each hidden layer containing $50$ neurons (and the learning rate is set as $10^{-5}$)\footnote{We have tried the same DQN structure as that in Section VII, but it does not perform well. One intuition is that the parameters in the two-hidden layer DQN with each layer containing $200$ neurons DQN is very large, which may require careful and longer training. Additionally, the two-hidden layer neural network may also not be able to provide a good approximation of Q values in more complex problems. Therefore, we decide to add one more hidden layer and reduce the number of neurons to $50$. This deeper DQN with fewer neurons has the ability to approximate more complicated Q-value function, and in the meanwhile requires less time to train before finding a good policy.}, and considered more complex simulated situations as well as real data traces. 

In this section, in addition to the Whittle Index heuristic, we also compare DQN with a Random Policy in which the user randomly selects one channel with equal probability at each time slot. Since the optimal policy even with a full knowledge of the system statistics is computationally prohibitive to obtain (by solving the Bellman-Ford equation in the belief state space) in general, we implement the Myopic policy as it is simple, robust and can achieve an optimal performance in some situations. However, one cannot consider the Myopic policy in general when system statistics is unknown since a single user is not able to observe the states of all channels at the same time so that one could not provide an estimation of the transition matrix of the entire system. Moreover, even if we allow the user to observe the states of all channels, the state space of the full system is too large to estimate and one would easily run out of memory when storing such a large transition matrix. Therefore, in the following simulation, we only consider cases when $\mathbf{P}$ is sparse and easy to access, and implement the Myopic policy as a genie (knowing the system statistics \emph{a-priori}) and evaluate its performance.

\subsection{Perfectly correlated scenario}

We consider a highly correlated scenario. 
In a $16$-channel system, we assume only two or three channels are independent, and other channels are exactly identical or opposite to one of these independent channels. This is the case when some channels are perfectly correlated, i.e., the correlation coefficient $\rho$ is either $1$ or $-1$. 

During the simulation, we arbitrarily set the independent channels to follow the same $2$-state Markov chain with $p_{11} \geq p_{01}$. When the correlation coefficient $\rho=1$, the user can ignore those channels that are perfectly correlated with independent channels and only select a channel from the independent channels. In this case, the multichannel access problem becomes selecting one channel from several i.i.d. channels that are positively correlated, i.e., $p_{11} \geq p_{01}$. Then as it is shown in the previous work~\cite{myopic_1, myopic_n}, the Myopic policy with known $\mathbf{P}$ is optimal and has a simple round-robin structure alternating among independent channels. In the case when $\rho=-1$, the Myopic policy with known $\mathbf{P}$ also has a simple structure that alternates between two negatively perfectly correlated channels. Though more analysis needs to be done in future to show whether the Myopic policy is optimal/near-optimal when $\rho=-1$, it can still serve as a performance benchmark as the Myopic policy is obtained with full knowledge of the system dynamics.

\begin{figure}
\vspace{-1cm}
\centering
\begin{minipage}{.45\textwidth}
  \centering
  \includegraphics[width=1\linewidth]{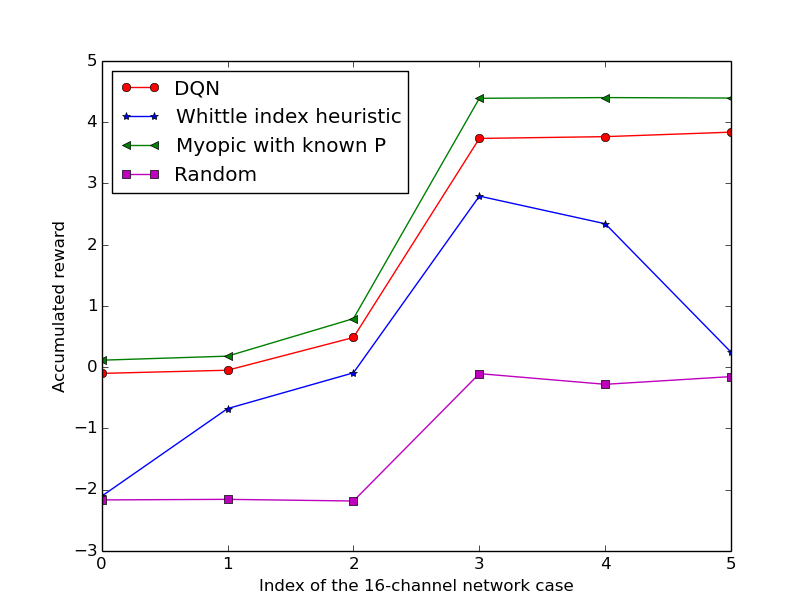}
  \captionof{figure}{Average discounted reward for 6 different cases. Each case considers a different set of correlated channels}
  \label{fig:16_reward}
\end{minipage}%
\hfill
\begin{minipage}{.45\textwidth}
  \centering
  \includegraphics[width=1\linewidth]{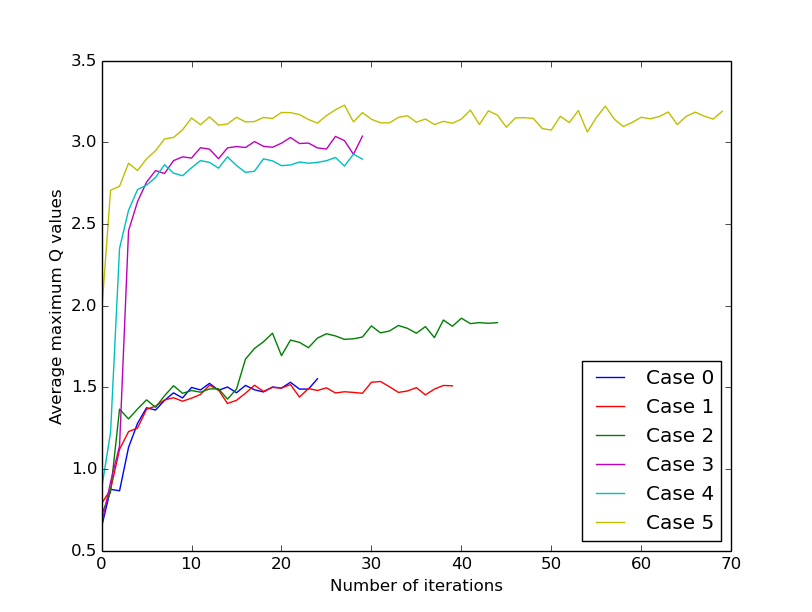}
  \captionof{figure}{Average maximum Q-value of a set of randomly selected states in 6 different simulation cases}
  \label{fig:q_value_fig}
\end{minipage}
\vspace{-0.5cm}
\end{figure}


In Fig.~\ref{fig:16_reward} we present the performance of all four policies: (i) DQN, (ii) Random, (iii) Whittle Index heuristic, and (iv) Myopic policy with known $\mathbf{P}$. In the first three cases (x-axis 0, 1 and 2), the correlation coefficient $\rho$ is fixed as $1$ and in the last three cases (x-axis 3, 4 and 5), $\rho$ is fixed as $-1$.
We also vary the set of correlated channels to make cases different. 
The Myopic policy in the first three cases is optimal, and in the last three cases is conjectured to be near-optimal. As it is shown in Fig.~\ref{fig:16_reward}, the Myopic policy, which is implemented based on the full knowledge of the system, is the best among all six cases and serves as an upper bound. 
DQN provides a performance very close to the Myopic policy without any knowledge of the system dynamics. The Whittle Index policy performs worse than DQN in all cases.

In addition, we collect the Q-values predicted from the DQN to show that DQN, indeed, tries to learn and improve its performance. 
Given a state $\mathbf{x}$, the maximum Q-value over all actions, i.e., $\max_{a} Q(\mathbf{x}, a)$, represents the estimate of the maximum expected accumulated discounted reward starting from $\mathbf{x}$ over an infinite time horizon. 
For each simulation case, we fix a set of states that are randomly selected, and then plot the average maximum Q value of all these states as the training is executed. 
As it is shown in Fig.~\ref{fig:q_value_fig}, in all cases, the average maximum Q-value first increases and then becomes stable, which indicates DQN learns from experience to improve its performance and converges to a good policy. As the environment cases are different, DQN may take a different amount of time to find a good policy, which is indicated as the different number of training iterations in each case in the figure for Q values becoming stable.


\subsection{Real data trace}
We use real data trace collected from our indoor testbed Tutornet\footnote{More information about the testbed on http://anrg.usc.edu/www/tutornet/} to train and evaluate the performance of DQN on real systems.
The testbed is composed of TelosB nodes with IEEE 802.15.4 radio.
We programmed a pair of motes distanced approximately 20~meters to be transmitter/receiver.
The transmitter continually transmits one packet to each one of the $16$ available channels periodically and the receiver records the successful and failed attempts. The transmitter switches transmitting on different channels so fast that the time difference can be ignored and the channel states of $16$ channels measured at each period can be considered to be in the same time slot. 
Both nodes are synchronized to avoid packet loss due to frequency mismatch and the other motes on the testbed are not in use.
The only interference suffered is from surrounding Wi-Fi networks and multi-path fading.
There are 8 Wi-Fi access points on the same floor and dozens of people working in the environment, which creates a very dynamic scenario for multichannel access.

The data are collected for around $17$ hours. 
Due to the configuration of Wi-Fi central channels, there are $8$ channels whose conditions are significantly better than others. 
Randomly selecting one channel from these good channels and keeping using it can lead to a good performance. 
Thus, in order to create a more adverse scenario and test the learning capability of the DQN, we ignore all these good channels and only use the data trace from the rest $8$ channels. 

We use the same data trace to train the DQN and to compute the MLE of the transition matrices of each channel for the Whittle index based heuristic policy. 
We compare the performance of the DQN policy, the Whittle index based heuristic policy and the Random policy. The Myopic Policy is not considered as finding the transmission matrix of the entire system is computationally expensive. The average accumulated discounted reward from each policy is listed in Table~\ref{tab:performance}. It can be seen that DQN performs best in this complicated real scenario. We also present the channel utilization of each policy in Fig.~\ref{fig:chann_util} to illustrate the difference among them. It shows DQN benefits from using other channels when the two best channels (used by the Whittle Index heuristic all the time) may not be in good states.

\begin{table}[ht]
\vspace{-0.5cm}
\begin{minipage}[b]{0.45\linewidth}
\centering
\vspace{-5cm}
\caption{Performance on Real Data Trace}
\begin{tabular}{cp{0.7\textwidth}}\hline
Method & \qquad Accumulated Discounted Reward \\
\hline
DQN & \qquad \qquad  $0.9473$\\
Whittle Index & \qquad \qquad $0.7673$\\
Random Policy & \qquad \qquad $-2.1697$\\
\hline
 \end{tabular}
    \label{tab:performance}
\end{minipage}\hfill
\begin{minipage}[b]{0.45\linewidth}
\centering
\includegraphics[width=1\textwidth]{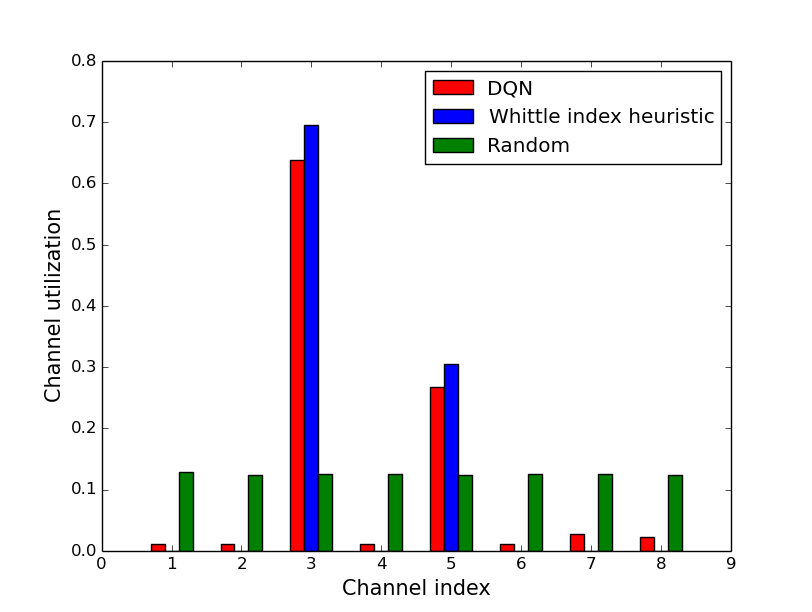}
\captionof{figure}{Channel utilization of 8 channels in the testbed}
\label{fig:chann_util}
\end{minipage}
\vspace{-0.5cm}
\end{table}

\subsection{Practical Issues}
From the previous analysis and experiment results, DQN shows a promising performance in the multichannel access problem. However, there are issues that need to be considered when implementing it in real deployments. In this paper when discussing the channel access problem, we only focus on one user and simply assume the user can always observe the actual state of his selected channel at each time slot. In practice, there are two entities involved, the sender and the receiver. They must be synchronized and use the same channel to communicate all the time. In a time slot when the sender selects a channel to transmit a packet, the receiver knows the selected channel condition based on whether it receives the packet or not, and the sender knows the selected channel condition from the acknowledgement (ACK) or negative-acknowledgement (NAK) message sent back by the receiver. If the receiver successfully receives the packet, it knows the channel is good and sends back an ACK to the sender, so that based on the received ACK the sender also knows the channel is good; if the receiver does not receive any packet, it knows the channel is bad and sends back an NAK to the sender, and thus the sender also knows the channel is bad. Therefore, when applying DQN in practice, we need to make sure the sender and the receiver always select the same channel at each time slot to guarantee their communication as well as having the same information about channel conditions through ACKs and NAKs.

One approach is to run the same structured DQNs at the sender and the receiver separately. The two DQNs start with the same default channel and are trained concurrently. We need to make sure the two DQNs have the same trained parameters and select the same channels at all times during training. Even though the ACK/NAK method can guarantee the sender and receiver have the same channel observations and thus training samples, there are still two facts that may cause the channel selection at the sender and the receiver to be different. First, in the exploration step, since each DNQ randomly selects a channel, it may happen that the two DQNs select different channels. Second, in the back propagation step, each DQN randomly selects a set of data samples from its experience replay to update its parameters. This may cause the parameters of two DQNs to become different,  which further results in different channel selection policies. To resolve the possible mismatch, we can use the same random seed on both sides to initialize the pseudorandom number generator in the implementation. In this way, the two DQNs always select the same random channel during exploration and use the same set of data samples to update parameters. Therefore, we can ensure the two DQNs will always select the same channel and the final learned policy is guaranteed to be the same. 

The channel mismatch problem can still happen when an ACK or NAK is lost (due to noise and/or interference) so that the sender and receiver might have different observations on the selected channel condition, and thus they may select different channels later. This inconsistent channel observation not only causes loss of communication, but also results in different learned DQN models at the sender and receiver that give different channel selection policies. One possible approach is to find a way to let the sender and the receiver be aware of the time when a channel mismatch happens, and try to recover in time. Since the sender is expecting to receive an ACK or NAK after each message is sent, the sender can detect the mismatch events if no ACK or NAK are received. Once the sender detects the possible channel mismatch event, it stops updating its DQN model as well as training dataset and transmits data in the future using one single channel - or a small set of channels known so far to have better channel conditions~\cite{kim2017fastjoining}. In addition, along with the original data messages, the sender also sends the timestamp when the channel mismatch was perceived. The sender keeps sending this channel mismatch time information until an ACK being received, which indicates the receiver is on the same channel again and receives the channel mismatch information. Therefore, the receiver can set its DQN model as well as its observation training dataset back to the state right before the channel mismatch happened (assume the receiver uses additional memory to store  different states of trained parameters and data samples), which guarantees that the sender and the receiver have the same DQN models as well as training datasets. They can resume operating and training thereafter. Suppose the sender only uses one current best channel to send the channel mismatch timestamp, and let $p_{good}$ be the probability of this channel being good in a time slot, $p_{ack}$ be the probability an ACK or NAK being lost, and $N$ be the number of channels in the system. As the receiver keeps training its DQN model before being aware of the channel mismatch problem, it applies the $\epsilon$-greedy exploration policy (explained in Section VII-A) during training phase. Therefore, with probability $\epsilon$, the receiver randomly picks a channel. Thus, after a channel mismatch happens, the probability that the sender and the receiver meet again on the same good channel and at the same time the ACK is successfully received is $\frac{\epsilon p_{good}(1-p_{ack})}{N}$. Once they meet on the same good channel, they can re-synchronize. Based on the above approach, the expected number of time slots required for re-syncing after a channel mismatch is $\frac{N}{\epsilon p_{good}(1-p_{ack})}$. Since the ACK packet is very small, the probability of loss is small~\cite{decouto2005highthroughput}. As long as the sender and the receiver can re-synchronize again after a channel mismatch, the effectiveness of the proposed policy is guaranteed and the performance will not be affected too much on average.

\section{Adaptive DQN for Unknown, Time-Varying Environments}
\label{sec:adaptive-dqn}
The studies in previous sections all focus on stationary situations, and DQN performs well in learning and finding good or even optimal dynamic multichannel access policies. However, in practice, real systems are often dynamic 
across time, and our DQN framework in previous sections cannot perform well in such situations. This is because we keep evaluating the newly-learned policy after each training iteration and once a good policy is learned\footnote{In this paper, we manually check the evaluation performance and stop the learning when a policy is good enough. More advanced techniques such as Secretary Problem~\cite{secretary} (by considering each learned policy as a secretary) can be used to decide when to accept a policy and stop learning.}, our DQN framework stops learning and keeps following this good policy. Thus, it lacks the ability to discover the change and re-learn if needed. To make DQN more applicable in 
realistic situations, we have designed an adaptive algorithm in Algorithm~\ref{alg:adaptiveDQN} to make DQN able to be aware of the system change and re-learn if needed. The main idea is to let DQN periodically evaluate the performance (i.e., the accumulated reward) of its current policy, and if the performance degrades by a certain amount, the DQN can infer that the environment has changed and start re-learning.

On the other hand, the Whittle Index heuristic cannot detect the environment change by simply observing the reward change. This is because the policy given from Whittle Index heuristic is far from the optimal policy, and it may have the low performance in both old and new environments so that there is no significant change in the reward leading to the claim that the environment has changed. In addition, even if the Whittle Index heuristic could detect the change, the new policy may still give a bad performance as the Whittle Index heuristic ignores the correlations among channels and is not able to have a correct estimation of the system dynamics due to its limited partial observation ability.

\begin{algorithm}[!ht]
    \small
  \caption{\small{Adaptive DQN}}\label{alg:adaptiveDQN}
  \begin{algorithmic}[1]
  \State First train DQN to find a good policy to operate with
  \For{$n=1,2,\ldots$ }
  \State At the beginning of period $n$
  \State Evaluate the accumulated reward of the current policy 
  \If{The reward is reduced by a given threshold\footnotemark}
  \State Re-train the DQN to find a new good policy
  \Else
  \State Keep using the current policy
  \EndIf
  \EndFor
  \end{algorithmic}
\end{algorithm}

\footnotetext{The threshold is set by the user according to her preference.}

\begin{figure}
\vspace{-1cm}
\centering
\begin{minipage}{.45\textwidth}
  \centering
  \includegraphics[width=1\linewidth]{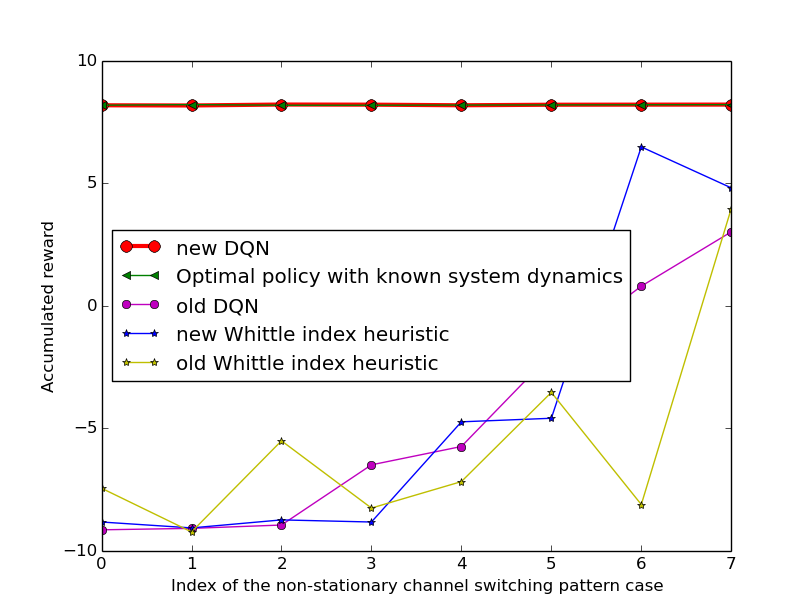}
  \captionof{figure}{Average discounted reward as we vary the 
  channel switching pattern situations}
  \label{fig:pattern_change}
\end{minipage}%
\hfill
\begin{minipage}{.45\textwidth}
  \centering
  \includegraphics[width=1\linewidth]{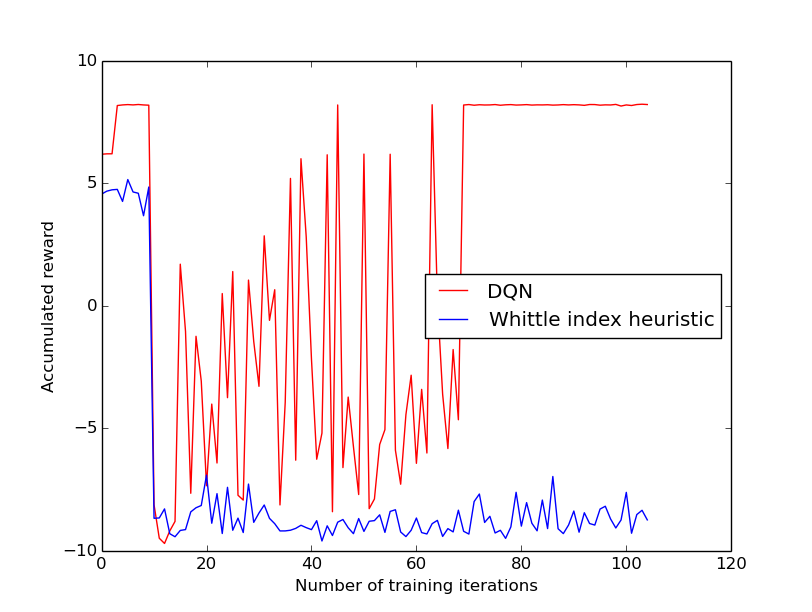}
  \captionof{figure}{Average discounted reward in real time during training in 
  unknown fixed-pattern channel switching}
  \label{fig:realtime}
\end{minipage}
\vspace{-0.5cm}
\end{figure}

In the experiment, we make the system initially follow one of the fixed-pattern channel switching cases from Section~\ref{sec:simulation-deterministic-switching}, and after some time it changes to another case. We consider both single good channel and multiple good channel situations. We let DQN automatically operate according to Alg.~\ref{alg:adaptiveDQN}, while we manually re-train Whittle Index heuristic when there is a change in the environment. Fig.~\ref{fig:pattern_change} compares the reward of both the old and new policies learned for DQN and the Whittle Index heuristic in the new environment, as we vary the pattern changes. As can be seen, DQN is able to find an optimal policy for the new environment as the genie optimal policy does, while Whittle Index heuristic does not. 

We also provide the real-time accumulated reward during the learning process of DQN and the Whittle Index heuristic in one of the above pattern changing situations in Fig.~\ref{fig:realtime}. The system initially starts with an environment that has $8$ channels being good at each time slot for the first $10$ iterations. As can be seen, both DQN and the Whittle Index heuristic are able to quickly find a good channel access policy, but DQN achieves the optimal performance. At iteration $11$, the environment changes to having only $1$ channel being good at each time slot. As there is a significant drop in the reward, DQN can detect the change and starts re-learning. And at iteration $70$, DQN finds the optimal policy and our system keeps following the optimal policy thereafter. On the other hand, even though we manually enable the Whittle Index heuristic to detect the change and re-estimate the system model and re-find a new policy, its performance is still unsatisfying as it cannot make use of the correlation among channels.

\section{Conclusion and Future Work}
\label{sec:conclusion}

In this paper, we have considered the dynamic multichannel access problem in a more general and practical scenario when channels are correlated and system statistics is unknown. As the problem, in general, is an unknown POMDP without any tractable solution, we have applied an end-to-end DQN approach that directly utilizes historical observations and actions to find the access policy via online learning. In the fixed-pattern channel switching, we have been able to analytically find the optimal access policy that is achieved by a genie with known system statistics and full observation ability. Through simulations, we have shown DQN is able to achieve the same optimal performance even without knowing any system statistics.
We have re-tuned the DQN implementation, and shown from both simulations and real data trace that DQN can achieve near-optimal performance in more complex scenarios. 
In addition, we have also designed an adaptive DQN and shown from numerical simulations that it is able to detect system changes and re-learn in non-stationary dynamic environments to provide a good performance. 

\begin{figure}[]
    \vspace{-1cm}
    \centering
  \includegraphics[width=.5\textwidth]{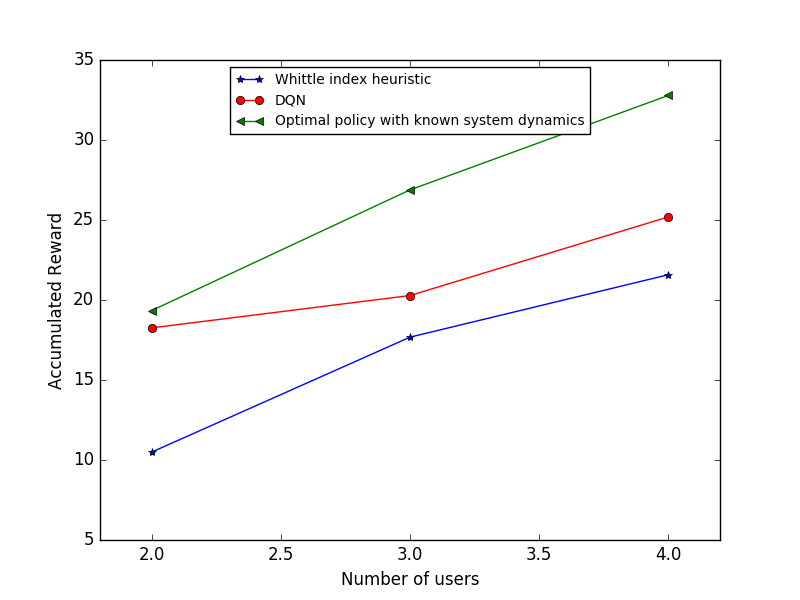}
    \caption{Average discounted reward as we vary the number of users in the multiple-user situation}
    \label{fig:multi_user}
    \vspace{-0.5cm}
\end{figure}

There are a number of open directions suggested by the present work.  First, we plan to apply the DQN framework to consider more realistic and complicated scenarios such as multi-user, multi-hop and simultaneous transmissions in WSNs. The framework of DQN can be directly extended to consider these practical factors in a simple way. For example, in the situation of multiple users, to avoid interference and collisions among users, we can adopt a centralized approach: assuming there is a centralized controller that can select a subset of non-interfering channels at any time slot, and assign one to each user to avoid a collision. By redefining the action as selecting a subset of non-interfering channels, the DQN framework can be directly used for this multi-user scenario. As the action space becomes large when selecting multiple channels, the current DQN structure requires careful re-design and may require very long training interval before finding a reasonable solution. Instead, we use the same DQN structure as that in Section VII and consider the multiple-users situation in a smaller system that contains $8$ channels where at any time slot $6$ channels become good and channel conditions change in a round-robin pattern. The number of users varies from $2$ to  $4$. As is shown in Fig.~\ref{fig:multi_user}, DQN can still achieve a good performance in the multiple-user case. Other deep reinforcement learning approaches, such as Deep Deterministic Policy Gradient (DDPG)~\cite{ddpg}, will be studied in future to tackle the large action space challenge.

Second, when the number of users in the network becomes large, the above proposed centralized approach becomes too computationally expensive to implement in practice. In future, we plan to study a more practical distributed approach where each user can learn a channel selection policy independently. One intuitive idea is to implement a DQN at each user independently. Then users can learn their channel selection policies parallelly, and avoid interference and conflicts by making proper channel-selection decisions based on the information gained from observations and rewards. However, whether a good or optimal policy can be learned, and whether an equilibrium exists are unknown and need further investigation.

Moreover, as DQN is not easy to tune and may get stuck in local optima easily, we plan to spend more time improving our DQN implementation as well as considering other Deep Reinforcement Learning approaches to see if they have the ability to reach the optimal performance in general situations and study the tradeoff between implementation complexity and performance guarantee. Also as a way to test the full potential of DQN (or Adaptive DQN) as well as other deep reinforcement learning technologies in the problem of multichannel access, we encourage the networking community to work together to create an open source dataset (like what has been done in computer vision and NLP community) that contains different practical channel access scenarios so that researchers can benchmark the performance of different approaches. We have published all the channel access environments and real data trace considered in this work online\footnote{https://github.com/ANRGUSC/MultichannelDQN-channelModel}. This might serve as an initial dataset for researchers to use.

\bibliographystyle{IEEEtran}


\end{document}